# An EUV Non-Linear Optics Based Approach to Study the Photochemical Processes of Titan's Atmosphere

*Jérémy Bourgalais[1], Nathalie Carrasco[1], Ludovic Vettier[1], Thomas Gautier[1], Valérie Blanchet[2], Stéphane Petit[2], Dominique Descamps[2]. Nikita Fedorov[2], Romain Delos[2], Jérôme Gaudin[2]*

[1]LATMOS-IPSL, Université Versailles St-Quentin, CNRS/INSU, Sorbonne Université, UPMC Univ. Paris 06, 11 boulevard d'Alembert, 78280 Guyancourt, France (e-mail: jeremy.bourgalais@latmos.ipsl.fr)

[2]CELIA, Université de Bordeaux – CNRS – CEA, UMR5107, 351 Cours de la Libération F33405 Talence, France

**ABSTRACT**

In situ exploration of the planetary atmospheres requires the development of laboratory experiments to understand the molecular growth pathways initiated by photochemistry in the upper layers of the atmospheres. Major species and dominant reaction pathways are used to feed chemical network models that reproduce the chemical and physical processes of these complex environments. Energetic UV photons initiate a very efficient chemistry by forming reactive species in the ionospheres of planets and satellites. Here we present a laboratory experiment based on a new photoreactor with an irradiation beam produced by high order harmonic generation of a near infra-red femtosecond laser. This type of EUV source is nowadays stable enough to enable long-lasting experiments during which a plethora of individual reactions can take place. Its high accessibility is such that chemical initial conditions can be systematically varied to investigate the complexity of the upper atmosphere of planets. In order to demonstrate the validity of our approach, we shone during 7 hours at 14 eV with a flux of $10^{10}$ photons sec$^{-1}$ cm$^{-2}$, a $N_2$/$CH_4$ (5%) based gas mixture defined by a 60 µm free mean-path. Such conditions are able to mimic the photochemistry of Titan $N_2$ upper atmosphere. The reaction products reveal the formation of hydrocarbons and N-bearing species like dimethyldiazene ($C_2H_6N_2$), the largest compound detected in this new photoreactor. This work represents an important step in the use of a EUV irradiated closed-cell chamber for the generation of photochemical analogues of Titan aerosols to better constrain the growth pathways of nitrogen incorporation into organic aerosols in the Titan atmosphere.

# 1. INTRODUCTION

## 1.1. Photochemistry of Titan's atmosphere

For the last two decades, advances in observational techniques have revolutionized our knowledge about the Universe and it does not end there. The arrival of a new generation of telescopes on the ground and in space (e.g., JWST, TESS, WFIRST, ELTs)[1–4] allows us, each year, to look further afield in the sky and through spectroscopic observations, to identify and quantify detected molecules in many astrophysical environments (e.g., molecular clouds, circumstellar envelopes, starburst galaxies, planetary atmospheres). Those analyses are non-trivial and in order to solve the conundrums arising from data collected by space missions and earth-based facilities, or either to prepare future spacecraft missions, laboratory experiments have to be well-honed to provide accurate information in many different physical conditions. This is the main challenge in order to propose comprehensive/credible models of the formation and destruction mechanisms of the molecules in many astrophysical environments.

Among the plenty intriguing objects within the solar system, Titan, the biggest moon of Saturn, is unique as a proxy to the early Earth. Titan has a thick atmosphere containing significant quantities of nitrogen and carbon through its main components ($N_2$ and $CH_4$)[5], liquid hydrocarbons lakes on the surface[6] along with hydrological activities.[7] The exposition of Titan's highest layer (> 700 km) to external sources of energy (mainly solar photons and electrons from the Saturnian magnetosphere) leads to the photoionization and photodissociation of the most abundant molecular species. Those processes trigger an efficient photochemistry forming small (< 100 amu) complex molecules (*e.g.,* nitriles, hydrocarbons) including both neutral and ionic species for which most of observed abundances are now reproduced reasonably well via photochemical models.[8–18] Those molecules induce a progressive formation of more complex organic compounds (*e.g.,* polycyclic aromatic hydrocarbons, polycyclic aromatic nitrogen heterocycles) at lower altitudes (ca. 500 km)[19] which are the building block toward the formation of aerosols (sub-µm particles)[20] forming haze layers in the atmosphere and falling down to the surface of Titan due to gravity.

This complex chemical activity triggered several past (Pioneer 11, Voyager I & II, Cassini-Huygens)[21,22] and future (Dragonfly)[23] spacecraft missions to Titan and many laboratory experiments during the last two decades. However, in spite of all the efforts to understand the molecular growth pathways building up complexity in the atmosphere of Titan, large areas of concern remain with unanswered questions especially regarding the formation of heavier molecules, preventing us to grasp a complete picture of Titan's planet evolution. (see Hörst 2017[24] for a recent review of our current understanding of Titan's atmosphere).

*1.2. Extreme ultra-violet UV irradiation experiments*

Since current *in situ* and direct observations of Titan are insufficient to understand Titan's atmosphere chemistry, the last decades have witnessed the rise of experimental laboratory simulation of Titan's atmosphere. (see Coll et al. 2013[25] for a review) The global idea is to simulate the atmosphere by mimicking the initial step, *ie.* exciting a $N_2$/$CH_4$ (5%) gas mixture with different sources of energy (electrons, ions, protons, photons, X- and gamma-rays) by using plasmas or light sources, in order to produce Titan's aerosol analogues and investigate both the gas and solid phase (Titan's aerosol analogues so-called Tholins) products to understand the whole cycle. However, the most efficient processes to activate the chemistry based on $N_2$ and $CH_4$ results from the interactions with solar photons.[26] While the less energetic (UV range) photons penetrating deeper in the atmosphere are absorbed by secondary products as $C_2H_2$ or $C_4H_2$, the high energetic photons (Extreme Ultra Violet (EUV) range) are absorbed mainly by molecular nitrogen in the upper layer of the atmosphere. The interesting photon range [80-100] nm couples methane and nitrogen chemistry. It lies above the methane dissociation energy (98 nm ca. 12.6 eV) and below the molecular nitrogen ionization threshold (79.4 nm ca. 15.6 eV) leading to both ground state and electronic excited states atomic N-fragments for which their role is still largely unknown.

So far the usual EUV sources employed for experimental laboratory simulations of Titan's atmosphere are low-pressure mercury and deuterium lamps limited to producing photons of 185 and 254 nm with ca. $10^{16}$ photons s$^{-1}$ cm$^{-2}$.[27–32] With such EUV sources, only the photodissociation process of methane can be reached and the typical radiation exposure time is up to hundreds of hours to produce an analyzable quantity of aerosols[33–35] due to weak photochemical solid products yield.[36] This yield that results from the slow kinetic reaction rate at that low photon energies, is typically three orders of magnitude lower than the plasma-generated aerosols.[37,38] For such long lasting experiments with an unavoidable carbon deposition on optical surfaces, cautions should be thus observed concerning the stability of the photon flux. With optical transmission being affected with the exposure time, the photochemistry is then altered within the reaction cell as the efficiency of the chemical growth depends on the relative amounts of photons to reactants and the optical depth of the gas mixture in the chamber.[39]

So far, the most appropriate EUV source that is stable, intense ($10^{12-15}$ ph s$^{-1}$ cm$^{-2}$) and tunable is synchrotron radiation. These large synchrotron facilities allow to investigate photochemical aerosols analogues.[40–42] Their unique disadvantage is the restricted access (ca. only few days/year) that by limiting the duration for the experimental campaigns reduces the physical multi-parametric studies. An alternative to synchrotron radiation is the use of microwave plasma discharge EUV lamps, which are made of a surfatron-type RF discharge using a rare gas (helium, neon or argon) flow in the mbar pressure range coupled to a photochemical reactor.[43] However despite an efficient high (up to $10^{14}$ ph s$^{-1}$ cm$^{-2}$) flux of EUV photons, they emit on specific atomic lines (He I: 58.43 and 53.70 nm, Ne I: 73.59, 73.37 nm and Ar: I 104.8, 106.6 nm) and no emission lines lie in the narrow range of interest for $N_2$ only photodissociation, ie. 98-79 nm. Another alternative light source is the high harmonic generation (HHG) of femtosecond laser.[44,45] These light sources deliver typically 200 meV broadband pulses at fixed wavelengths corresponding to odd harmonics of the fundamental laser. This fundamental wavelength can be tuned[46] to get large coarse tunability. This table-top source emits from the extreme UV down to the water window spectral range[47], depending on the generation gas chosen and the laser parameters. Combined with monochromator, selective dielectric mirror, or either metallic filter as used in the present work, such monochromatic EUV emission can be used as a possible, and way more accessible, alternative to synchrotron radiation sources. In the present work, we demonstrate the feasibility on the photochemistry of Titan.

*1.3. Our approach*

In this work, we have developed a new photo-reactor, SURFACAT, able to investigate the photochemical processes of Titan's atmosphere at 13.9 eV (the 9$^{th}$ harmonics - H9 of a femtosecond laser centered at 800nm-89.2nm). The aim was to provide a beam stable over several hours with the highest photon flux (ca. $10^{10}$ photons s$^{-1}$ cm$^{-2}$) in a closed cell for the first time in order to irradiate a $N_2$/$CH_4$ (5%) gas mixture in the reactor chamber during variable irradiation times. The neutral species produced in the photochemical cell were monitored by a quadrupole mass spectrometer allowing to measure ppm species.

## 2. RESULTS

Molecular species from the first photochemical experiments mimicking the chemistry in Titan's upper atmosphere by using a EUV-HHG beamline are displayed in **Figure 1**. The reactive gas mixture in the closed cell was irradiated for 7 hours at 14 eV with a measured photon flux of ca. $2 \times 10^{10}$ photons $s^{-1}$ $cm^{-2}$. The mass spectrum was measured at the end of the experiment after release of the condensed photoproducts from the cryogenic trap (see Methods). The experiment was reproduced twice ensuring reproducibility.

The mass spectrum in **Figure 1** is displayed over the 2-80 mass range as no significant signal is observed at larger masses. The mass spectrometer used in this work used an electron ionization source to ionize sample gases coming from the reactor prior to their detection. Molecules interact with an electron beam with a typical 70 eV energy, which initiates molecular fragmentation and yields several peaks at different m/z resulting in fragmentation patterns (FPs). The measured mass spectrum thus corresponds to the addition of multiple FPs of the molecules formed in the reactor, which requires a decomposition to address the contribution of each molecule. This decomposition can be restrained by the limited knowledge one can have of molecule fragmentation patterns in the ionization source of the instrument used. To overcome this limitation we used a novel approach to deconvolution, recently developed by Gautier et al.[48], where a Monte-Carlo sampling is applied to the expected FPs. This allows for a better retrieval of the sample composition and its constituent mixing ratios. The interpretation of peaks below m/z 25 is not considered in this retrieval; being too complex due to the residual signature of water in the mass spectrometer along with the signature of fragments of larger molecules.

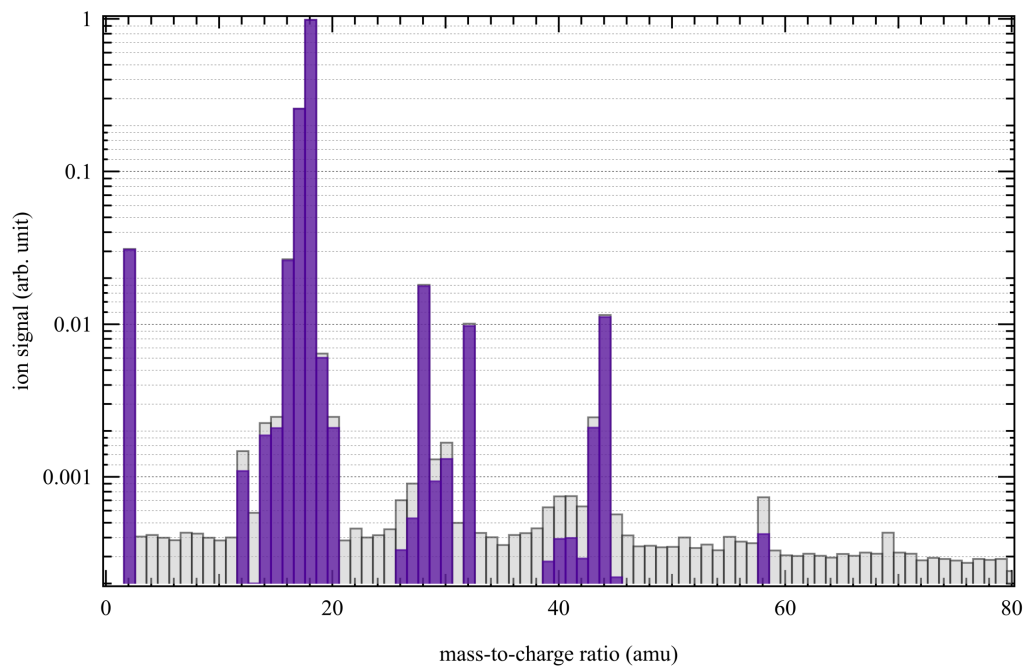

**Figure 1.** Mass spectrum with (purple bars) and without (grey bars) background subtraction obtained after 7 hours of trapping time at 13.9 eV.

The database given in the **Table S1** of the Supplementary Material summarizes the species used to perform deconvolution and contains the FPs of the species which are available on the NIST database. The result of the deconvolution is displayed in **Figure 2**. The purpose of the present discussion is not to be exhaustive in all the detected mass peaks but to highlight the main neutral photoproducts and compare them to Titan photochemistry.

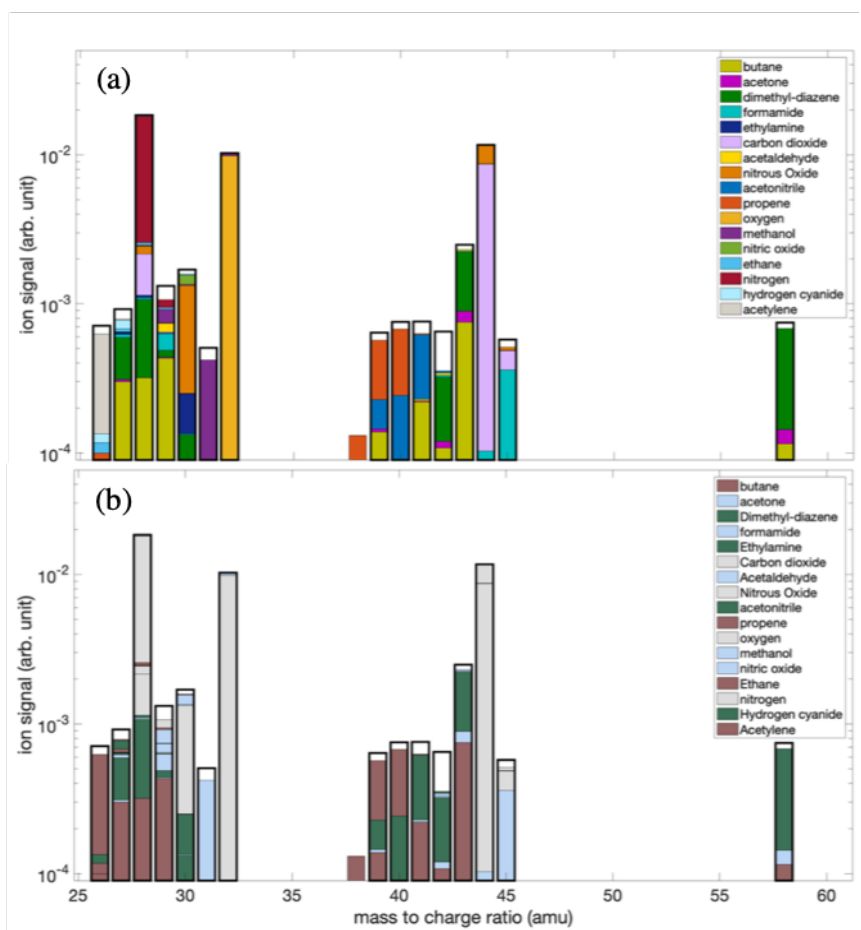

**Figure 2. (a)** Decomposition of the mass spectrum shown in **Figure 1** using the Monte-Carlo approach from Gautier et al.[48]. The experimental mass spectrum used for the decomposition is shown with the bold black lines. The colored bars show the calculated contribution of each of the molecules to the mixture. (b) The colored bars show the contribution of different families of chemical compounds: residual gas reactants (grey), oxygenated compounds (blue), hydrocarbons (brown) and N-bearing species (green).

As visible in **Figure 2**, our algorithm is able to reconstruct the mass spectrum with minor exceptions. In order to evaluate the quality of the reconstruction of the experimental spectrum, **Figure 3** displays the percentage error between the result of the decomposition of the mass spectrum and the experimental mass spectrum. All peaks are reproduced with a relative error of less than 20% except for mass 42 which exceeds 40%. Such good agreement below 20% for almost a full mass spectrum is rarely achieved and is consistent with the instrumental variability of the measured fragmentation patterns compared to the reference NIST database. The database accurately reproduces the experimental spectrum and determines the contribution of each molecule referenced in the analyzed gas mixture. The relative error of mass 42 although superior to the others appears reasonable in the eyes of errors on FPs. However, it cannot be discarded that a molecule or a fragment is missing in the database at this mass 42 .

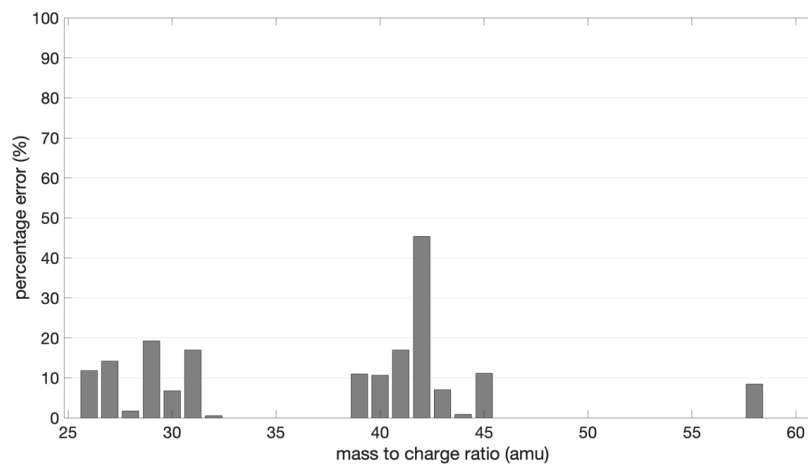

**Figure 3.** Relative error of deconvolution with the experimental mass spectrum for each mass channel.

*Contaminants and oxygenated species:*

When looking into detail the molecular attributions obtained from the deconvolution process, we observe dominant signals at m/z 28, 32 and 44 assigned mainly at nitrogen $N_2$, oxygen $O_2$, nitrous oxide $N_2O$ and carbon dioxide $CO_2$ respectively. These ion signals are due to a residual gas phase at the top of the cryogenic trap coming from a slow but inevitable accumulation of air in the closed cell during the 7 hours through the micro-leaks of the reactor. The introduction of oxygen in the experiment involves undesirable partial oxidation of the organic compounds in the experiment as can be seen with the formation of nitric oxide NO, methanol, formamide and acetone at m/z 30, 32, 45 and 58 respectively. However despite the formation of oxygenated species and contaminants, the deconvolution allows the identification of major hydrocarbons and N-species (cf. **Figure 2b**). One should stress that these oxygenated do not act as chemical intermediates in the formation of the oxygen-free products we will now focus on.

*Hydrocarbons:*

Fragments of acetylene $C_2H_2$ and ethane $C_2H_6$, which are abundant products in the chemistry of Titan[49] are detected at m/z 26, 27, 28, 29 and 30 respectively, comforting the ability of the experiment to trigger Titan's complex chemical network starting from methane and nitrogen photolysis. Other light hydrocarbon, such as ethene $C_2H_4$ observed in Titan's atmosphere cannot be characterized in the present experiment even if produced as it is too volatile to be directly condensed in the cryogenic trap. More relevant to demonstrate that HHG offers an interesting alternative to synchrotron radiation, is the formation of heavier hydrocarbons as the $C_3$-hycrocarbon propene $C_3H_6$ at m/z 42 which was observed in Titan's stratosphere by Nixon et al. (2013)[50] and $C_4$-hycrocarbon butane $C_4H_{10}$ at m/z 58.[51]

*N-species:*

The smallest N-species detected in this work is hydrogen cyanide at m/z 27 which is a detected well-known stable molecule[52] and acting as a brick to build up more complex nitrile species[53] as for instance acetonitrile $C_2H_3N$ at m/z 41. This well-known product[54] is easily formed through the rearrangement of product isomers from the reaction between excited nitrogen $N(^2D)$ coming from the photolysis of nitrogen in the reactor and ethene $C_2H_4$.[55] However most of the nitrile gas photochemistry is still to investigate. Ethylamine $C_2NH_5NH_2$ (mass 45) with as main signature a peak at mass 30, was found as the most abundant amine in nearly all Titan aerosol analogues studies in the literature[32] and was a typical product detected in glow discharge generated in $N_2/CH_4$ gas mixture.[56] The detection of ethylamine in this work supports the assumption over its abundance in Titan's aerosols, although this compound is still not yet be included in Titan atmospheric models of aerosol formation. Finally the detection of dimethyldiazene $C_2H_6N_2$ (CH3-N=N-CH3) at m/z 58 is consistent with the large nitrogen incorporation observed by the aerosols collector pyrolysis instrument of the Huygens probe.[57] This is the largest mass produced in the present experiment after 7 hours of irradiation at 14 eV and is a very encouraging result on the path of formation of photochemical aerosol analogues in the laboratory.

3. **DISCUSSION AND PERSPECTIVES**

First results were obtained after 7 hours of irradiation at 14 eV with an estimated flux of ca. $2 \times 10^{10}$ photons s$^{-1}$ cm$^{-2}$. Despite the precautions taken, the analysis of the mass spectrum highlights the presence of oxygenated species (NO, $O_2$, $CO_2$, $CH_3OH$) coming from the slow but inevitable micro-leaks with the reactor in closed cell.

Most important, 60% of the products that have been detected are due to photochemistry only based on initial gas mixture ($N_2$ and $CH_4$). We see the formation of small hydrocarbons such as acetylene, ethane and propene. While acetylene is one of the major compounds in Titan's atmosphere, we also detected heavier hydrocarbons propene and butane which are both predicted or observed in Titan's chemistry. Additionally, this work addresses the formation of N-rich organic products (hydrogen cyanide, acetonitrile, ethylamine, dimethyldiazene) even if there is a big lack in the NIST database for such fragmentation pattern. Among them, several nitrile compounds such as HCN and acetonitrile $CH_3CN$ have been firmly detected and quantified in Titan's atmosphere. Other heavier N-compounds produced in this work have been detected in Titan aerosol analogues studies in the literature including the most abundant amine, ethylamine.[32]

Our results highlight the formation of an efficient N-rich photochemistry at 13.9 eV close to Titan's atmospheric chemistry. This work represents a significant step using a closed cell chamber for the generation of photochemical Titan aerosol analogues in order to better constrain the nitrogen fixation processes in Titan's atmosphere and to its relevance to the evolution of early life.

This project is the first milestone of a long-term strategy to exploit the potential of HHG as EUV source during planetary investigations. Present limitations due to contamination with the slow but inevitable micro-leaks which progressively introduce oxygen in the reactor will require improvements such as the use of higher EUV flux to decrease the irradiation time. This new type high repetition rate EUV sources (above 200kHz) have been recently optimized[46] paving the way to shorter irradiation time, and consequently leak free experiments.

## 4. METHOD

The experimental setup can be sub-divided in 2 main parts: the high harmonic generation (HHG) beamline, including generation and beam transport and the atmospheric chamber for photochemical tholins generation (SURFACAT) and diagnostics as depicted in **Figure 5** of the entire experimental setup.

### *4.1. High-harmonic generation*

We used the AURORE femtosecond Ti:sapphire laser at CELIA laboratory in Bordeaux (France) which is feeding up to 5 five beamlines dedicated to ultra-fast phenomena. The laser is delivering 7 mJ pulses of 30 fs FWHM duration at a repetition rate of 1 kHz. The spectrum is centered at a wavelength of 800 nm (hv = 1.55 eV) with a FWHM of 50 nm. The laser beam is focused by a plano-convex lens of 1.5 m focal in a gas jet backed by $C_2H_2$ gas. The typical backing pressure was 130 mbar, while the pressure in the generation chamber is rising up to $10^{-2}$ mbar. A differential pumping stage is located in between the generation chamber and the rejection chamber decreasing the pressure down to 7.2x10$^{-6}$ mbar. The beam can be either steered to the SURFACAT reactor by reflection on a SiO$_2$ mirror with a Nb$_2$O$_5$ anti-reflective coating for the 1.55 eV/800 nm (fundamental wavelength), or to the EUV spectrometer if the mirror is retracted. The reflectivity of this SiO2 mirror at 70° in the 14 eV range is maximized for a S polarization and has been measured to be 78.9%, while restraining the reflectivity of the 800 nm to 0.25%. These reflectivity values maintain the EUV flux without to melt the 210 nm-mesh supported Indium filter with the impinging intense 800 nm pulse. The In filter is the key optical element that both selects the 14 eV, defines the optical entrance of the reactor SURFACAT and maintains the volume/pressure of the reactants. The 14 eV radiation is characterized with the EUV spectrometer, made of entrance slit (1 mm aperture) and a variable line-space grating. The dispersed HHG beam as shown on **Figure 4** is then focused in the dispersion plane to a micro-channel plate which amplifies and converts the photon signal to an electron signal.

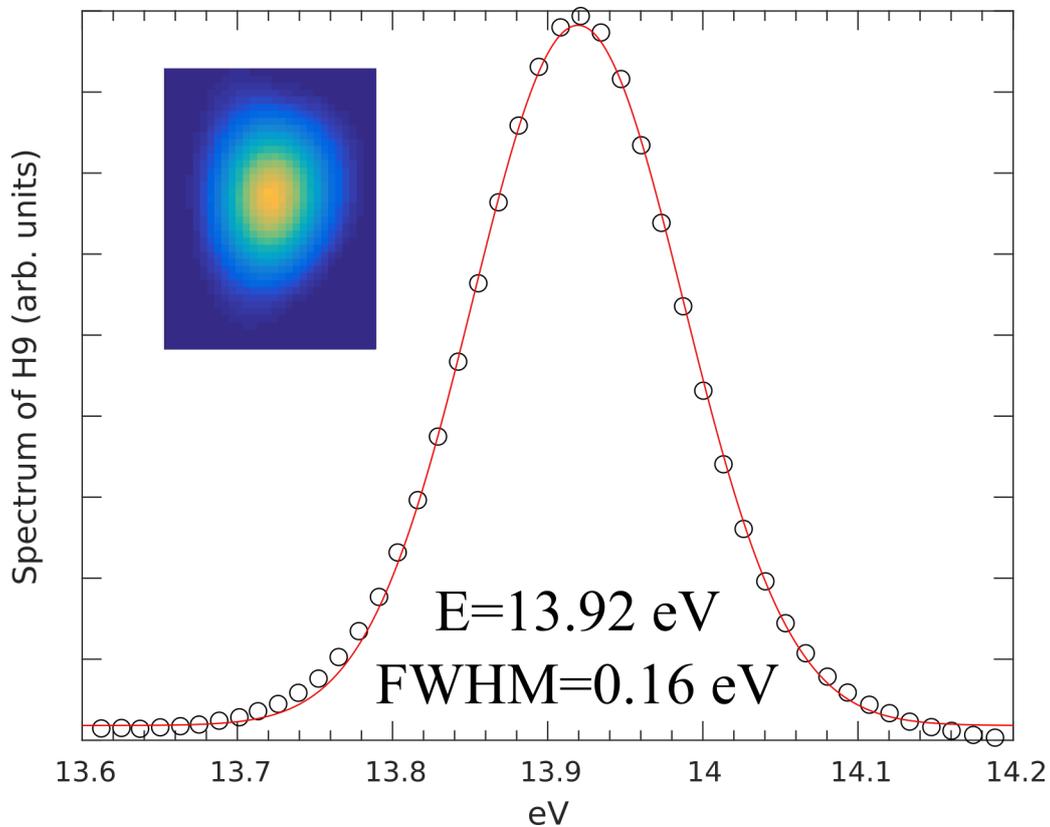

**Figure 4**: Spectrum of H9 centered at 13.9 eV recorded on the VUV spectrometer with its spatial vertical mode in inset (horizontal mode is the plane of dispersion of the grating).

The electron signal is then converted back to visible photons signal which can be monitored by a CCD camera located outside the vacuum chamber. This spectrometer allows us to monitor the HHG to tune online the different relevant parameters to optimize the H9 photon flux. The full width half maximum spectral bandwidth of the H9 was measured to be 0.16 eV. As the beam is refocused only along the dispersion plane the size of the measured spot in the vertical direction corresponds to the beam size, allowing us to measure the beam divergence. This later one is measured to be $9 \times 10^{-2}$ degree. The distance from the HHG source - Indium filter is fixed such that the HHG beam diameter is equal to the clear aperture of the filter, i.e. 15.9 mm. Once the optimization is achieved the rejection mirror is inserted in the beam. The 210 nm thick In filter acts like a monochromator. In fact it has a transmission of $1.7 \times 10^{-2}$ (H7 = 10.8 eV), 0.21 (H9 = 13.9 eV) and $4.6 \times 10^{-3}$ (H11 = 17.0 eV) (CXRO data http://henke.lbl.gov/optical_constants/). We have measured a transmission of 15% at 14 eV. The photon flux measured with a 100 mm$^2$ XUV photodiode (Opto Diode AXUV 100G) was 8 nA corresponding to a flux of ca. $2 \times 10^{10}$ ph/s.

## *4.2. The atmospheric chamber (SURFACAT) coupled to a cryogenic trap*

At the opposite side, the reactor is closed with a glass window to check for light alignment. During all the irradiation time, this window is overshadowed.

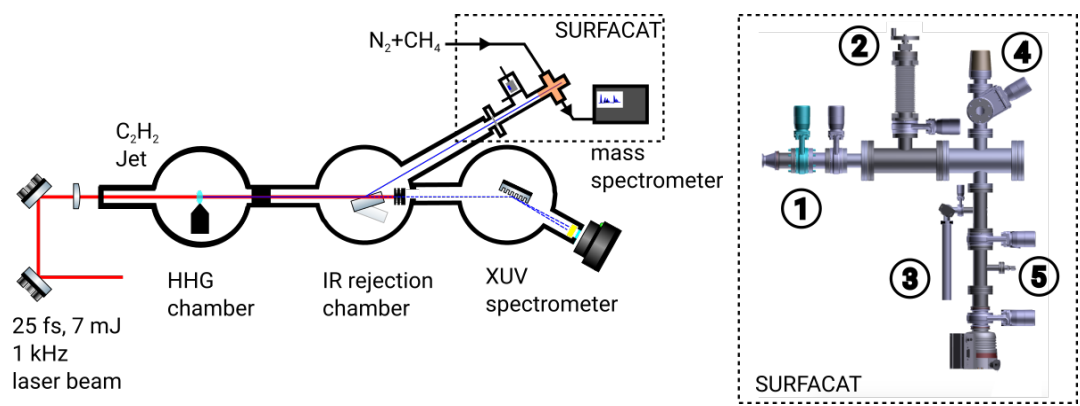

**Figure 5.** Schematic diagram of the HHG beamline and the SURFACAT setup: 1) In filter, 2) retractable XUV photodiode, 3) cold trap, 4) entrance of the mass spectrometer, 5) gas inlet

The chamber for photochemical tholins generation, is a CF-63 cross piece stainless steel reactor, presented in **Figure 5**, where a Titan's atmosphere relevant $N_2/CH_4$ (5%) gas mixture is introduced up to 1 mbar total pressure by using a 10 sccm (standard centimeter cube per minute) range flow controller from the underside. The typical mean-free path is then around 60 μm. One fundamental aspect is the contamination from the wall reactor as well as possible leak. As a consequence prior to the experiment the reactor was backed up to 120 °C during 70h. The pressure dwindled to ~$10^{-6}$ mbar. We then tested possible leak by switch off the pumping. After 10 days the pressure was at $3.5 \times 10^{-5}$ mbar. Before EUV irradiation, all CF parts were then again backed up to 120°C for 9 hours.

The pressure is monitored with an absolute capacitance gauge and, before each experiment, the reactor is pumped down to ~$10^{-6}$ mbar by a primary and a turbo molecular pump located in the lower part, in order to clean out the chamber from residual gas traces. During the photochemistry experiments, removable VAT vacuum valves from each side isolate the reactor and ensure a stable pressure on the order of 1 mbar during few hours of irradiation.

In order, to maximize the detection of the products during our experiment, a cryogenic trap held at liquid nitrogen temperature (77 K at atmospheric pressure) was used to capture *in situ* at the end of the irradiation time and accumulate the compounds for an efficient *ex-situ* analysis using the MS. The cold trap is positioned at the closest spot from the MS and the setup was minimized as possible to increase the density of products. At the end of each experiment, few hours of irradiation, the valve towards the cold trap at liquid nitrogen temperature is opened, enabling to accumulate the condensable gas-phase products during few minutes in the trap. At this temperature molecular nitrogen and most of $C_2$-hydrocarbons are not trapped efficiently.[58] Then, the trap is isolated and warmed up to room temperature during half hour and open on the SURFACAT reactor. The volatile products are released under vacuum and mass spectrometry analysis is performed.

### 4.3. *Mass spectrometry diagnostic*

Finally, the top side of the cross piece is connected to a mass spectrometer (HIDEN Analytical HPR-20 QIC) to monitor the neutral gaseous products (1 % to 0.01 % range) during variable irradiation times in the experiment. In the MS, neutral molecules are ionized by a 70 eV electronic ionization and detected with a resolution of 1 atomic mass unit (u) and over a 100 u mass range. Gas sampling is achieved through a metal-bellow tube radially close to the irradiated chamber (cf. **Figure 5**), also ensuring a relatively low enough pressure (< $10^{-5}$ mbar) in the MS during the sampling. The **Figure 1** displays a mass spectrum when the MS was connected to the chamber under vacuum (ca. $10^{-6}$ mbar).

### 4.4. *Deconvolution of neutral mass spectrum*

Upon ionization in a RGA, neutral species tend to undergo ionizing dissociation, leading to the formation of a specific fragmentation pattern for each species. We used these fragmentation patter to deconvolve the mass spectra and retrieve the individual contribution of each species present in the reactor following the method described in detail in Gautier et al.[48]. This method assumes that the measured mass spectra is a linear combination of each species concentration multiplied by their fragmentation patterns. This is true if the only source of ion in the mass spectrometer is from neutral-electron interaction in the ion source, which is the case in nominal pressure condition for laboratory RGA. We briefly remind here the principle of this method:

A mass spectrum is decomposed (*i.e.* to retrieve the species relative concentrations) into individual species contribution using interior-point least square fitting on the suspected species fragmentation patterns. These fragmentation patterns are obtained from databases such as the NIST, but are highly dependent of the geometry of the ionization source of the instrument for the measurement. This means that typically, the fragments intensity for a given species measured in a lab can vary by up to 50% compared to databases, rendering decomposition doubtful. The method used here allow for compensating this issue by using a Monte-Carlo sampling of the fragmentation patterns, and performing the mass spectra deconvolution not once but several thousands of times with as many different fragmentation pattern. This allows for the retrieval of the probability densities. The retrieved composition is then corrected to account for the ionization cross section of each compounds to ultimately retrieve the mixing ratio of each individual compound detected in the mass spectrum.

## 5. AKNOWLEDGEMENTS

This research was supported by the ERC Starting Grant PRIMCHEM, grant agreement n°636829. This work was partially funded by the IdEx Bordeaux Programm of the Agence Nationale de la Recherche through the LAPHIA (ANR-10-IDEX-03-02).